\documentclass[12pt]{article} \input epsf.tex
\usepackage{epsfig,graphicx,psfrag,axodraw}
\usepackage{amsmath,amsfonts,amssymb,stmaryrd,latexsym}
\usepackage{url}
\newcommand{\be}{\begin{equation}}
\newcommand{\ee}{\end{equation}}
\newcommand{\bea}{\begin{eqnarray}}
\newcommand{\eea}{\end{eqnarray}}
\newcommand{\bear}{\begin{eqnarray}}
\newcommand{\eear}{\end{eqnarray}}

\newcommand{\ba}{\begin{array}}
\newcommand{\ea}{\end{array}}

\newcommand{\cc}{C}\newcommand{\ac}{a}\newcommand{\mc}{{\cal P}}

\begin{document}

\baselineskip=18pt \pagestyle{plain} \setcounter{page}{1}

\vspace*{-1cm}

\noindent \makebox[11.6cm][l]{\small \hspace*{-.2cm}  October 11, 2012}{\small FERMILAB-Pub-12-557-T}  \\  [-1mm]
\makebox[11cm][l]{\small \hspace*{-.2cm}  Revised: February 26, 2013}{\small } \\ [-2mm]

\begin{center}

{\Large \bf   Coupling spans of the Higgs-like boson   } \\ [9mm]

{\normalsize \bf Bogdan A. Dobrescu and Joseph D. Lykken\\ [3mm]
{\small {\it
Theoretical Physics Department, Fermilab, Batavia, IL 60510, USA }}\\
}

\end{center}


\begin{abstract}
Using the LHC and Tevatron data, we set upper and lower limits on the total width of the
Higgs-like boson. The upper limit is based on the well-motivated assumption that the Higgs coupling to a $W$ or $Z$ pair 
is not much larger than in the Standard Model. These width limits allow us to convert the 
rate measurements into ranges for the Higgs couplings to various particles. 
A corollary of the upper limit on the total width is an upper limit on the branching fraction of exotic Higgs decays. 
Currently, this limit is  47\% at the 95\% CL if the electroweak symmetry is broken only by doublets.
\end{abstract}



\section{Introduction} \setcounter{equation}{0}

The discovery of a Higgs-like particle ($h^0$) in the $\gamma\gamma$ and $4\ell$ final states by the ATLAS \cite{:2012gk} and CMS \cite{:2012gu} Collaborations,
of mass $M_h$ roughly in the $125-126.5$ GeV range,
provides multiple opportunities for probing phenomena beyond the Standard Model (SM).
The SM Higgs boson \cite{Djouadi:2005gi} in that mass range has a very small total width, $\Gamma_h^{\rm SM}/M_h \simeq 3.2 \times 10^{-5}$,
due to the small Yukawa coupling of the $b$ quark ($y_b \sim 0.02$) and the severe phase-space suppression of the $WW^*$ final state.
Therefore, if new particles lighter than about 60 GeV have a coupling to the Higgs doublet larger than $10^{-2}$, the Higgs-like boson can have a large
branching fraction ${\cal B}_X$ into exotic final states, and consequently a larger total width, $\Gamma_h > \Gamma_h^{\rm SM}$.
The exotic Higgs decays could escape detection for a long time, for example in the case of the four gluon-jet final state arising from
 $h\to A^0A^0 \to 4 g$  where $A^0$ is a gauge-singlet  spin-0 particle  \cite{Dobrescu:2000jt}.

Thus, it is important to analyze whether a relatively large $\Gamma_h$ can be observable. 
The prospects for measuring the line shape of $h^0$ are rather dim, barring a high-luminosity muon collider
running at $\sqrt{s} = M_h$.  Nevertheless, 
one may hope to determine $\Gamma_h$ indirectly, given that all the rates 
for Higgs signals at colliders are inversely proportional to $\Gamma_h$. 
It turns out, however, that the effect on rates of a larger $\Gamma_h$ can be compensated by an universal increase of the $h^0$ couplings.
In fact, at hadron colliders the only observables based on rates are a product of the squared couplings for producing and decaying $h^0$ divided by $\Gamma_h$,
so that the width itself cannot be measured even indirectly at the LHC.\footnote{At $e^+e^-$ or $\mu^+\mu^-$ colliders the recoil of the $Z$ produced in association with $h^0$ 
would allow a measurement of the $h^0ZZ$ coupling, and consequently $\Gamma_h$ can be determined from the rates for various processes.}

The impossibility of measuring  $\Gamma_h$ at the LHC  hampers the extraction of  Higgs couplings from the rate measurements. 
In order to go around this problem, several groups have relied on assumptions about the width. 
For example, it has been assumed 
that only SM decays are allowed \cite{Zeppenfeld:2000td, Duhrssen:2004uu, Carmi:2012yp}, 
or that all nonstandard final states include particles escaping the detector \cite{Peskin:2012we, Carmi:2012in, Englert:2012wf},
or that nonstandard final states are allowed only when certain couplings to SM particles are the same as in the SM
\cite{Lafaye:2009vr}.   A more general framework is allowed in 
\cite{Banerjee:2012xc}, but the problem of rescaling both the width and the couplings is avoided by imposing an ad-hoc upper limit on some of the couplings.

In this paper we use the ATLAS and CMS rate measurements to derive an upper limit on $\Gamma_h$ based on 
the rather robust theoretical assumption that the Higgs coupling to a $W$ pair is not much larger than in the SM.\footnote{This procedure was mentioned in Refs.~\cite{Duhrssen:2004uu, Peskin:2012we, LHCHiggsCrossSectionWorkingGroup:2012nn}
but was not explored in detail in the general case where all couplings are free parameters and nonstandard decay modes are allowed.}
In many models,  the $h^0WW$  coupling is 
substantially smaller than in the SM, and only in the unusual case \cite{Georgi:1985nv,Logan:2010en,Low:2010jp,Falkowski:2012vh}
of large VEVs for higher $SU(2)_W$ representations 
does the coupling exceed its SM value.\footnote{An  $h^0WW$ coupling larger than in the SM also leads to unitarity violation in longitudinal $WW$ scattering unless there 
are isospin-2 resonances \cite{Falkowski:2012vh}.}
We then translate this limit on $\Gamma_h$ into an upper limit on the exotic branching fraction, ${\cal B}_X$. 
Furthermore, we derive a lower limit on $\Gamma_h$ from the Tevatron \cite{Aaltonen:2012qt}
and LHC \cite{:2012gk, :2012gu} rate measurements, especially for the $b\bar{b}$ and $WW^*$ decay modes.
Having bounded  $\Gamma_h$ from above and below, we can then extract nearly-model independent 
upper and lower limits on the general couplings of the Higgs-like particle allowed by various rate measurements.
Our method of deriving the spans of the couplings could be used by the CMS and ATLAS Collaborations in order to translate their measurements into 
information about the couplings of the Higgs-like particle in a way that is as model-independent as possible at hadron colliders.

In Section 2 we parametrize the general couplings of the Higgs-like boson to SM particles, and summarize the 
existing rate measurements.  The upper and lower limits on $\Gamma_h$ are derived in Section 3. 
In Section 4 we compute the upper limit on the branching fraction of exotic Higgs decays, and then we obtain the coupling spans.
Our conclusions are included in Section 5.

\section{Measurable quantities}\setcounter{equation}{0}

A Higgs boson is a scalar particle $h^0$ that couples to the 
$W$ and $Z$ bosons according to 
\be
\frac{g}{M_W} \, h^0 \left( \cc_W M_W^2 W^+W^- + \cc_Z \, \frac{M_Z^2}{2}  ZZ \right)  ~,
\label{coupling:W}
\ee
where $g$ is the $SU(2)_W$ gauge coupling,
and $\cc_W$ and $\cc_Z$ parametrize the deviation from the SM couplings: 
$\cc_W^{\rm SM}=\cc_Z^{\rm SM}=1$.
If electroweak symmetry breaking is due entirely to VEVs of 
$SU(2)_W$ doublets, then \cite{Pocsik:1982cz, Duhrssen:2004uu, Peskin:2012we}
\be
0 < \cc_W = \cc_Z \leq 1  \;\;\;\;  {\rm for \;\, doublet \;\, VEVs}  ~~.
\label{eq:constraint-doublet}
\ee

In models where  triplets or higher  $SU(2)_W$ representations acquire VEVs 
it is possible to have $\cc_W \neq \cc_Z$ as well as values for $\cc_W$ and/or 
$\cc_Z$ above 1 or negative \cite{Logan:2010en}. However, such models predict additional 
scalars, including doubly-charged and singly-charged particles, whose effects are 
tightly constrained by the electroweak data \cite{Haber:1999zh} and by collider searches \cite{Chiang:2012dk}. 
As a result, one can still derive some upper bounds on the couplings:
\be
|\cc_W| < \cc_W^{\rm max}  \;\;\; , \;\;\;\; |\cc_Z| < \cc_Z^{\rm max}   ~~.
\label{eq:constraint}
\ee
with  $\cc_W^{\rm max},  \cc_Z^{\rm max} = O(1)$.

For example, the Georgi-Machacek model \cite{Georgi:1985nv} includes a real 
triplet and a complex triplet (such that 
custodial invariance may arise due to a cancellation between the contributions of the two triplets),
and also a complex doublet whose VEV is necessary for giving the top mass.
Due to the loop contributions of charged scalars to the $Zb\bar{b}$ vertex \cite{Haber:1999zh},
the deviations from the SM couplings have upper limits $\cc_W^{\rm max}, \cc_Z^{\rm max} \approx 1.5$
\cite{Logan:2010en}. 

The couplings of a Higgs boson to third generation fermions can be written as
\be
- \cc_t \frac{m_t}{v_h} \, h^0  \, \bar{t} t
- \cc_b \frac{m_b}{v_h} \, h^0  \, \bar{b} b - \cc_\tau \frac{m_\tau}{v_h} \, h^0  \, \bar{\tau} \tau ~~,
\label{coupling:b}
\ee
where $m_t$, $m_b$ and $m_\tau$ are the $t, b$ and $\tau$ masses, 
$v_h \approx 246$ GeV is the electroweak scale, 
and $\cc_t, \cc_b, \cc_\tau$ are real parameters that are equal to 1 in the SM.

The Higgs boson coupling to the top quark, and possibly to new colored particles, 
induces a 1-loop coupling of $h^0$ to a pair of gluons.
In the approximation where $M_h/m_t$ effects 
are negligible (they turn out to be below 7\% for $M_h \approx 125$ GeV) and where new 
colored  particles that couple to $h^0$ can be integrated out, the Higgs coupling to a pair of gluons 
is given by a dimension-5 operator:
\be
\cc_g \, \frac{\alpha_s}{12 \, \pi v_h} \, h^0  \,G^{\mu\nu}G_{\mu\nu}   ~~,
\label{coupling:g}
\ee
where $G^{\mu\nu}$ is the gluon  field strength,
and $\cc_g$ is a real parameter (equal to 1 in the SM).
If there are new colored particles with mass not much larger than $M_h$ that couple to $h^0$ 
(such as a color-octet scalar \cite{Bai:2011aa}), then $\cc_g$ should be replaced by a function of $M_h$.

The only other Higgs couplings relevant here involve photons and arise also at one loop. 
These lead to the $h^0 \to \gamma\gamma, \, Z\gamma$ decays.  Given that the
dominant contributions to these decays in the SM arise from $W$ loops \cite{Djouadi:2005gi}, it is not accurate to parametrize these couplings
by the dimension-5 operators $h^0  \, F^{\mu\nu}F_{\mu\nu}$ and $h^0  \, Z^{\mu\nu}F_{\mu\nu}$.
For example, in the SM the full 1-loop $\Gamma^{\rm SM} (h^0 \to \gamma\gamma)$ 
width is 50\% larger than the result based on the dimension-5 operator for $M_H \approx 125$ GeV.
We are thus led to define the deviations from the SM effective couplings to photons by
\bear
&& \cc_\gamma \equiv \left(\frac{\Gamma (h^0 \to \gamma\gamma)}{\Gamma^{\rm SM} (h^0 \to \gamma\gamma)}\right)^{\! 1/2}  ~~,
\nonumber \\ [2mm]
&& \cc_{Z\gamma} \equiv \left(\frac{\Gamma (h^0 \to Z\gamma)}{\Gamma^{\rm SM} (h^0 \to Z\gamma)}\right)^{\! 1/2}   ~~.
\label{coupling:gamma}
\eear

There are several processes  at hadron colliders that can be studied in order to determine 
the couplings shown in Eqs.~(\ref{coupling:W}), (\ref{coupling:b}), (\ref{coupling:g}) and (\ref{coupling:gamma}).
Higgs production proceeds mainly through gluon fusion,
vector boson fusion (VBF), $Wh^0$ or $Zh^0$ associated production, or radiation off a top
quark ($t\bar{t}h^0$). 
The cross sections  for these five processes are proportional to $\cc_g^2$, $(\cc_W^2 + r \cc_Z^2)$,
$\cc_W^2$, $\cc_Z^2$, and $\cc_t^2$, respectively. The parameter $r$ that sets the ratio of rates for
$ZZ$ to $WW$ fusion is typically between 0.3 and 0.5 in $pp$ collisions, and depends 
on the center-of-mass energy \cite{Han:1992hr}.

The widths for the Higgs decays to $b\bar{b}$, $\tau^+\tau^-$, $WW$, $ZZ$, $\gamma\gamma$, $Z\gamma$,
are proportional to $\cc_\mc^2$ where $\mc =  b,\tau, W,Z,\gamma, Z \gamma$,
respectively. Additional decay modes, 
to final states involving SM particles ({\it e.g.}, $h^0 \to A^0A^0 \to 4j$ where $A^0$ is a new spin-0 particle \cite{Dobrescu:2000jt}), 
or new stable particles may have large contributions to  $\Gamma_h$. 

The narrow width approximation is accurate for $M_h \approx 125$ GeV 
even if new physics contributions to $\Gamma_h$ were three orders of magnitude larger
than $\Gamma_h^{\rm SM}$,  the total Higgs width in the SM. 
Thus, the cross section for a process of Higgs production and decay is proportional to 
\be
\frac{\cc_{\rm prod.}^2 \, \cc_{\rm decay}^2}{\Gamma_h}  ~~,
\label{sketch}
\ee
where $\cc_{\rm prod.}$ and $\cc_{\rm decay}$ are the $\cc_\mc$ 
coefficients entering the production and decay, respectively, as discussed above.
It is convenient to define the ``apparent squared-couplings" 
\be
\ac_\mc  \equiv \cc_\mc^2 \left(\frac{\Gamma_h^{\rm SM}}{\Gamma_h}\right)^{\! 1/2}  \;\;\; , \;\;\;\;
{\rm for} \;\, \mc = W,Z,g , \gamma, Z \gamma , t, b,\tau ~~,
\label{relation}
\ee
so that the cross section for a Higgs process [proportional to the quantity in Eq.~(\ref{sketch})] 
over the SM cross section for the same process is simply a product of two $\ac_\mc$  quantities.\footnote{In the notation of \cite{LHCHiggsCrossSectionWorkingGroup:2012nn}, 
our $C_\mc$ parameters are labelled by $\kappa_\mc$, 
and $a_\mc = \kappa_\mc^2/\kappa_H$.}

The measurements of various Higgs processes allow the determination of the $\ac_\mc$ apparent squared-couplings. 
For example, $\ac_b$ and $\ac_W$ may be extracted from the measured total cross sections for 
$pp \to W^* \to W h^0$ followed by $h^0 \to b\bar{b}$ or $h^0 \to  W^+W^-$, through the relations:
\bear
\left(\frac{\sigma}{\sigma_{\rm SM}}\right)
(Wh \to Wb\bar{b}) = a_W \, a_b    ~~,  \nonumber 
\\ [2mm]
\left(\frac{\sigma}{\sigma_{\rm SM}}\right)(Wh \to WWW) =  a_W^2  ~~,
\eear 
where $\sigma$ is the measured cross section and $\sigma_{\rm SM}$ is its theoretical value in the SM.
Likewise, measurements of the cross sections for various Higgs production mechanisms followed by 
various Higgs decays determine other products of $\ac_\mc$'s, as shown in Table 1.

A fit to the measured cross sections listed in Table 1 can determine $\ac_\mc$ for $\mc = b, W, Z, g, \tau, \gamma$.
Only channels that are already measured or will be probed in the near future are included in Table 1.
Many other channels 
such as $gg\to h^0 \to Z\gamma$ (proportional to $a_g \ac_{Z\gamma}$), $Wh^0$ production followed by $h^0 \to ZZ^*$ (proportional to $\ac_W \ac_Z$),  
 $Zh^0$ production followed by $h^0 \to \tau\tau$ (proportional to $\ac_Z \ac_\tau$),  
or $t\bar{t} h$  production followed by $h^0 \to W^+W^-, ZZ, \tau^+\tau^-$, will likely require a 
sizable integrated luminosity due to their small rates or large backgrounds. 

\renewcommand{\arraystretch}{1.5} 
\begin{table}[p]\centerline{
\begin{tabular}{|c|c|c|c|}
\hline
 $h^0$ decay  &  $h^0$  production & observable 
  & measured $\sigma/\sigma_{\rm SM}$ ;  \ $M_h = 125$ GeV
\\  \hline\hline
             &  $gg \to h^0$     &   $ \ac_g \ac_W $     &    \parbox{5.9cm}{ ~ \\  $1.3 \pm 0.5$,  ATLAS \cite{:2012gk}   ; 126 GeV  \\  [0.5mm]
                                         $0.6^{+0.5}_{-0.4} $ , CMS \cite{:2012gu}  \ ; \ 125.5 GeV \\[1.5mm]  $0.3^{+0.8}_{-0.3}$ , Tevatron \cite{:2012zzl} \\ our average: $0.9 \pm 0.4$ \\ [-2mm] ~ } \\ \cline{2-4}
\parbox{0.9cm}{ $W W^* $ \\ ~ }                           &  VBF              &   $ \! (\ac_W \! + r \ac_Z)/(1\!+\!r) \, a_W \! $        &   $0.3^{+1.5} _ {-1.6}$ , CMS \cite{CMS-combination}  \\ \cline{2-4}
                           &  $W^* \to W h^0$  &   $\ac_W^2$                &  $-2.9^{+3.2} _ {-2.9}$ , CMS \cite{CMS-combination}   \\ \cline{2-4}
                          &  $Z^* \to Z h^0$  &   $\ac_Z \ac_W$          &    \\ \hline
\hline
\parbox{.8cm}{ ~ \\ [12mm] $ZZ^*$ }    &  $gg \to h^0$        &   $ \ac_g \ac_Z$                    &   \parbox{5.7cm}{ ~ \\  $1.3^{+0.7}_{-0.5}$ ,  ATLAS \cite{ATLAS-ZZ} 
                                      \\ [1mm] 
                                         $0.7^{+0.5}_{-0.4}$ , CMS \cite{:2012gu}  \ ; \ 125.5 GeV  \\ [1mm] our average: $1.0^{+0.4}_{-0.3}$   
                                          \\ [-2mm] ~ }   \\ \cline{2-4}
                                    &  VBF                 &   $\! (\ac_W \! + r \ac_Z)/(1\!+\!r) \, a_Z \! $       &                                      \\ \hline
\hline
\parbox{0.5cm}{ ~ \\ [6mm] $\gamma\gamma$}  &   $gg \to h^0$   &     $ \ac_g \ac_\gamma$  &    \parbox{6.3cm}{ ~ \\  $1.7\pm 0.6$ ,  ATLAS \cite{ATLAS-gamma} ; 126.5 GeV  \\      
                                                       $1.4 \pm 0.6$ , CMS \cite{CMS-combination}            \\ [1mm]  
                                                       $3.6^{+3.0}_{-2.5} $ , Tevatron \cite{:2012zzl} \\ [1mm] our average: $1.6 \pm 0.4$  \\ 
                                               [-3mm]  ~ } \\ \cline{2-4}
                &       VBF        &   $\! (\ac_W \! + r \ac_Z)/(1\!+\!r) \, \ac_\gamma \! $  &   \parbox{6.3cm}{ ~ \\  $2.6\pm1.3$ ,  ATLAS \cite{ATLAS-gamma} ; 126.5 GeV \\      
                                                      $ 2.1^{+1.4}_{-1.1}$ , CMS \cite{CMS-combination}   \\ [1mm] our average: $2.3^{+1.0}_{-0.9}$  \\ [-3mm]  ~ }  \\ \hline
\hline
\parbox{.4cm}{ ~ \\ [13mm] $b\bar{b}$ \\ [-8mm]}   
                &  \parbox{2.2cm}{~ \\ [1mm] $W^* \to W h^0$\\ [-2mm]}  &   $\ac_W \ac_b $    &   \parbox{5.9cm}{ ~ \\ [-2mm] $0.5\pm 2.2$ ,  ATLAS \cite{:2012gk} ; 126 GeV  \\  [1mm]
                                         \ $\;\; 0.5^{+0.9}_{-0.8}$ , CMS \cite{CMS-combination}  \\ [1mm] \ $\;\; 2.0 \pm 0.7$, Tevatron  \cite{:2012zzl} \\ our average: $1.4\pm 0.6$ \\ [-15mm] }   \\ \cline{2-3}
                &  \parbox{2.cm}{~ \\ [1mm] $Z^* \to Z h^0$ \\ [-1mm]}   &   $\ac_Z \ac_b$          &    \\ \cline{2-4}
                &  $t\bar{t} h^0$  &   $\ac_t \ac_b$          &   $-0.8^{+2.1}_{-1.9}$ , CMS \cite{CMS-combination}  \\ \hline
\hline
                &  $gg \to h^0$    &     $ \ac_g \ac_\tau$    &   \parbox{5.7cm}{ ~ \\  $0.4^{+1.6}_{-2.0}$ ,  ATLAS \cite{:2012gk}  ; 126 GeV \\ [1mm]
                                                       $1.3\pm 1.1$ , CMS \cite{CMS-combination}   \\ [1mm] our average: $1.0\pm 0.9$  \\ [-3mm]  ~ } \\ \cline{2-4}
\parbox{.7cm}{ ~ \\ [-10mm] $\tau^+\tau^-$  \\ [-2mm]}   &       VBF        &    $ \! (\ac_W \! + r \ac_Z)/(1\!+\!r) \, \ac_\tau \! $    
 &  $-1.8^{+1.0}_{-0.9}$ , CMS \cite{CMS-combination}   \\ \cline{2-4}
                & $W^* \to W h^0$  &  $ \ac_W \ac_\tau$     &  $0.7^{+4.1}_{-3.2}$  , CMS \cite{CMS-combination}    \\ \hline
\end{tabular}
}
\label{tab:charges}
\caption{Combinations of parameters (3rd column) that can be extracted from cross section measurements of various processes.  Existing measurements for
$M_h = 125$ GeV (except where $M_h$ is explicitly specified) are shown in the last column. 
Our averages do not include any correlations, and are obtained by combining asymmetric errors as in \cite{Barlow}.}
\end{table}

The measurements on the rows labelled by $gg \to h^0$ are dominated by gluon fusion but also contain some 
contributions of order 10\% from VBF and from associated production with a $W$ or $Z$ decaying hadronically.
For simplicity we neglect those contributions.

The measurements on the rows labeled by VBF include two forward jets. The selections used make VBF the dominant 
production mechanism, but do not eliminate completely the gluon fusion mechanism with two additional jets (simulations within the SM 
show that this contamination is about 30\% \cite{:2012gu}). 
This effect will convolute the determination of the $\ac_g$, $\ac_W$,
and $\ac_Z$ parameters, but taking it into account is beyond the scope of this article.

\section{Limits on the total Higgs width}\setcounter{equation}{0}

In this section we derive the lower and upper limits on the total width $\Gamma_h$ of $h^0$. 

\subsection{Upper limit on $\Gamma_h$} 

The existence of a stringent upper limit  $\Gamma_h^{\rm max}$ (with  $\Gamma_h^{\rm max}  \ll M_h$) on the total $h^0$ width is not obvious. 
After all, the observable quantities  $\ac_\mc$
can be kept fixed when $\Gamma_h$ is increased by increasing all the couplings $\cc_\mc$. The reason that there is a useful upper limit stems from the fact that 
there are upper limits on $\cc_W$ and $\cc_Z$, which are related to the  $W$ and $Z$ masses.

Once the  $\ac_W$ and $\ac_Z$ quantities are extracted from a fit to the data, each of the upper limits on 
the Higgs couplings to $WW$ and $ZZ$ gives an upper limit on $\Gamma_h$. 
Using the upper limits on $\cc_W$ and $\cc_Z$ as parametrized  in Eq.~(\ref{eq:constraint}), 
we find that Eq.~(\ref{relation}) implies the following upper limit on the total $h^0$ width:
\be
\Gamma_h \leq \Gamma_h^{\rm max}  = {\rm Min} \left\{ \frac{(\cc_W^{\rm max})^4}{\ac_W^2} \, , \,
 \frac{(\cc_Z^{\rm max})^4}{\ac_Z^2} \right\}  \;  \Gamma_h^{\rm SM} ~~.
 \label{upper-limit}
\ee
Note that this upper limit relies solely on the existence of an upper limit on the Higgs couplings to the $W$ or $Z$;  for example, it 
allows any contribution to the width from exotic Higgs decays.

In the case where the electroweak symmetry is broken only by the VEVs of $SU(2)_W$ doublets
(which covers the majority of theories discussed in the literature), the upper limit takes the form 
\be
\Gamma_h \leq  \Gamma_h^{\rm max}  =  \frac{ \Gamma_h^{\rm SM}  }{ \ac_V^2 }   ~~,
 \label{upper-limit-doublet}
\ee
where $\ac_V$ is now obtained by the fit performed with the $\ac_W = \ac_Z\equiv \ac_V$ constraint.
Note that $\ac_V$ can be measured directly from VBF or associated $Vh^0$ production followed by 
$h^0\to WW$ or $ZZ$. The experimental uncertainties in these channels are too large for now,
so that we use a more indirect method for extracting $\ac_V$. 

Let us first combine the 
$gg \to h^0 \to WW^*$ and $gg \to h^0 \to ZZ^*$ rate measurements shown in Table 1, using the prescription of Ref.~\cite{Barlow}
for asymmetric errors: 
\be
(\sigma/\sigma_{\rm SM})(gg \to h \to VV^* )  = 0.96^{+0.27}_{-0.24}  ~,
\ee
where $V = W$ or $Z$.
For $C_W = C_Z$, the observable quantity $\ac_V $ can be extracted from the
current measurements of $\sigma/\sigma_{\rm SM}$ for $h^0$ production followed by decay into $VV^*$ and $\gamma\gamma$:
 \be
\ac_V^2 = (\sigma/\sigma_{\rm SM})(gg \to h \to VV^* )  
\frac{ (\sigma/\sigma_{\rm SM})({\rm VBF} \to hjj \to \gamma\gamma jj )}{ (\sigma/\sigma_{\rm SM})(gg \to h \to \gamma\gamma )}    ~ ~.
\label{av2}
\ee

We assume that the quoted experimental uncertainties correspond to  Gaussian distributions, or to bifurcated Gaussian 
distributions ({\it i.e.}, two half-Gaussians of same central value glued together) in the case of asymmetric errors.
It should be emphasized that this is only a rough approximation, which could be avoided once more information about  experimental uncertainties becomes available.
With the inputs from Table 1, and manipulating the distributions with Monte-Carlo simulations, we find
\be
\ac_V =  1.15^{+0.39}_{-0.29} ~ ~.
\label{aV}
\ee
This implies the following upper limit on the total width:
\be 
\Gamma_h \leq \Gamma_h^{\rm max} = 0.52^{+0.82}_{-0.10} \,\, \Gamma^{\rm SM}_h  ~~.
\label{upper}
\ee
Note that, assuming the constraint from  $SU(2)_W$ doublets  ($\cc_W = \cc_Z\equiv \cc_V \leq 1$),  $\Gamma_h^{\rm max}$ is a strict upper limit on $\Gamma_h$,
but the extracted value of $\Gamma_h^{\rm max}$ from current data has an uncertainty; this is indicated at the  
$1\sigma$ level in Eq.~(\ref{upper}). 

Even though we assumed that the experimental inputs are (bifurcated) Gaussian distributions, the values of $\Gamma_h^{\rm max}$ follow a distribution that is quite different 
from a bifurcated Gaussian, due to the operations on Gaussians with large variances shown in Eqs.~(\ref{upper-limit-doublet}) and (\ref{av2}). 
The  $\Gamma_h^{\rm max}$ distribution obtained from current data is shown in Fig. 1.
The 95\% CL interval for $\Gamma_h^{\rm max}/\Gamma_h^{\rm SM}$ is $0.26-3.56$. 

\begin{figure}[t]\center
\vspace*{-0.7cm}\hspace*{0.6cm}
\psfig{file=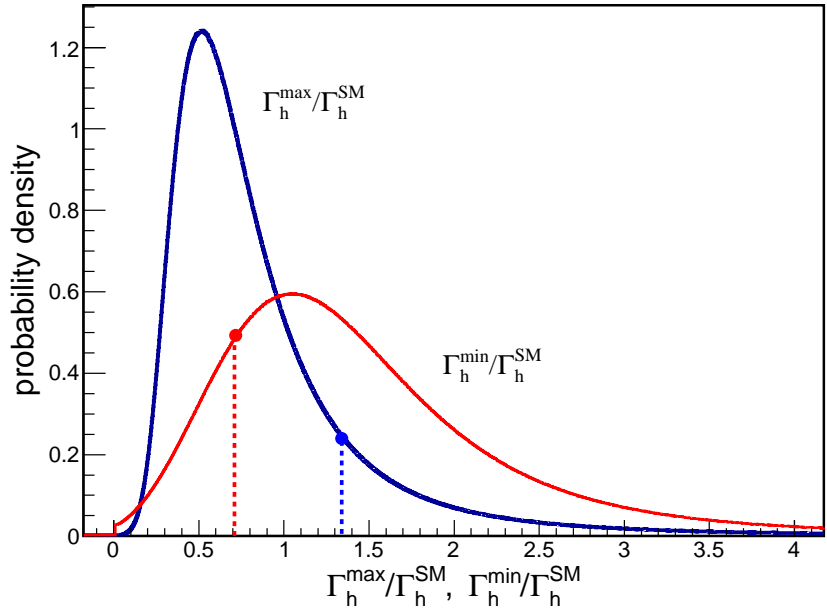 , width=12.3cm,angle=0}
\caption{ $\Gamma_h^{\rm max}$ distribution obtained from Eq.~(\ref{upper-limit-doublet}), and $\Gamma_h^{\rm min}$ distribution obtained from 
Eq.~(\ref{lower-limit}). 
The span of $\Gamma_h/\Gamma_h^{\rm SM}$ lies between the dashed vertical lines, which mark the lower  
$1\sigma$ limit on $\Gamma_h^{\rm min}$ and the upper $1\sigma$ limit on $\Gamma_h^{\rm max}$. 
}
\label{fig:GammaLimits}
\end{figure}

\subsection{Lower limit on $\Gamma_h$}

Unlike the above  upper limit which relies on a theoretical assumption (upper limit on the $hVV$ couplings), 
a lower limit on $\Gamma_h$ can be derived from the rates required for its observation.
The total width of the Higgs boson is given by
\be
\Gamma_h = \sum_{\parbox[t]{1.4cm}{\scriptsize $ \mc = W, Z, \\ \hspace*{0.3cm}   b, \tau, g,  \gamma $} } \!\! \cc_\mc^2 \; \Gamma^{\rm SM}( h^0 \! \to \mc\mc )  
+ \; \Gamma_X  ~~,
\label{total-width}
\ee
where $\Gamma_X$ is the $h^0$ decay width into final states other than the SM ones.
For simplicity, we have not included decays into $Z\gamma$, $c\bar{c}$ or 
light-fermion pairs because 
their sum is at most 3\% of $\Gamma_h$ in the SM for any $M_h > 120$ GeV.
In other words, decays into any of these final states with a width substantially larger than the SM one is 
included in the exotic width $\Gamma_X$.

Given that $\Gamma_X \ge 0$, Eq.~(\ref{total-width}) implies the following lower limit on the total Higgs width:
\be
\Gamma_h \ge \Gamma_h^{\rm min}  = \bigg(  \; \sum_{\parbox[t]{1.4cm}{\scriptsize $ \mc =  W, Z, \\ \hspace*{0.3cm}   b, \tau, g,  \gamma $} }\!\!
\ac_\mc \; 
{\cal B}^{\rm SM}( h^0 \!\to \mc\mc ) \bigg)^{\! 2} \;  \Gamma_h^{\rm SM} ~~,
\label{lower-limit}
\ee
where $ {\cal B}^{\rm SM}( h^0  \!\to \mc\mc )$ are the theoretically known branching fractions in the SM,
and  $\ac_\mc$ [defined in Eq.~(\ref{relation})] can be extracted from a fit to the rate measurements.
The fact that there is a lower limit on $\Gamma_h$ is not surprising given that the observation of a Higgs boson requires a sizable production rate 
which in turn requires couplings that are not much smaller than the SM ones. However, the exact form of the lower limit was
hard to anticipate.

We can extract the distribution for $\ac_b$ from the measurement of the rate for associated production followed by $h^0\to b\bar{b}$:
\bear
\ac_b &=& \frac{1}{\ac_V }  \, (\sigma/\sigma_{\rm SM})(V h^0\to Vb\bar{b} ) 
\nonumber \\ [2mm]
&=& 1.00^{+0.91}_{-0.37} ~~.   
\label{ab}
\eear
The distribution for $\ac_g$ can be obtained from the rate for gluon fusion followed by the $h^0\to WW^*, ZZ^*$ decays:
\bear
\ac_g &=& \frac{1}{\ac_V } \, (\sigma/\sigma_{\rm SM})(gg \to h^0 \to VV^* )  
\nonumber \\ [2mm]
&=& 0.74^{+0.35}_{-0.13}  ~~.
\eear
This then allows the determination of the remaining $a_\mc$ quantities:
\bear
\ac_\gamma &=& \frac{1}{\ac_g } \, (\sigma/\sigma_{\rm SM})(gg \to h^0 \to \gamma\gamma ) 
\nonumber \\ [2mm]
&=& 1.88^{+0.65}_{-0.46} ~~.
\\ [4mm]
\ac_\tau  &=&  \frac{1}{\ac_g } \, (\sigma/\sigma_{\rm SM})(gg \to h^0 \to \tau^+ \tau^-) 
\nonumber \\ [2mm]
&=& 1.0^{+1.5}_{-0.9} ~~.
\label{atau}
\eear

Note that $\ac_\tau$ could also be extracted from the rate for the VBF process  followed by $h^0 \to \tau^+ \tau^-$.
As can be seen from Table 1, the central value for this rate is about $2\sigma$ {\it below} the predicted value
for the case of no Higgs boson. This suggests a large negative fluctuation of the background, so we have chosen
not to include this VBF process in a fit until more data is analyzed.

The large uncertainty in $\ac_\tau$ shown in Eq.~(\ref{atau}) raises the issue of what is the meaning of a negative $a_\mc$.
Clearly, negative values for $\sigma/\sigma_{\rm SM}$ represent downward fluctuations of the background.
However,  $a_\mc$ are by definition [see Eq.(\ref{relation})]  positive quantities, so that it is appropriate to interpret 
the uncertainties quoted in Eqs.~(\ref{ab})-(\ref{atau}) as distributions with a boundary at the origin.
Following the Feldman-Cousins \cite{Feldman:1997qc} prescription for that case (and assuming approximate Gaussians with variance
given by the negative errors), the $1\sigma$   
confidence interval for $\ac_\tau$ becomes $0.3 - 2.5$. 
For the purpose of determining the lower limit on $\Gamma_h$, it does not make much difference whether we use 
this interval or the one indicated by Eq.~(\ref{atau}), $0.1 - 2.5$, because ${\cal B}^{\rm SM}( h^0 \!\to \tau^+\tau^- )$ is rather small. 

For $M_h = 125$ GeV, 
${\cal B}^{\rm SM}( h^0 \to \mc\mc)$ equals (21.5, 2.64, 57.7, 6.32, 8.57, 0.228)\%
for $\mc = W, Z,  b, \tau, g,  \gamma $, respectively, and the total SM width is 
$\Gamma_h^{\rm SM} = 4.07$ MeV.
The lower limit on $\Gamma_h$ can then be computed from Eq.~(\ref{lower-limit}):
\be
\Gamma_h \geq \Gamma_h^{\rm min} = 1.05^{+1.26}_{-0.34} \, \, \Gamma_h^{\rm SM}  ~~.
\label{lower}
\ee
We reiterate that $\Gamma_h^{\rm min}$ is a strict lower limit for  $\Gamma_h$, but that the value of  $\Gamma_h^{\rm min}$
extracted from current data has an uncertainty represented here by the 68.3\% CL range. The 95\% CL interval for  
$\Gamma_h^{\rm min}/\Gamma_h^{\rm SM}$ is $0.30 - 4.95$.

Using the upper limit at the 68.3\% CL for $\Gamma_h^{\rm max}$ given in Eq.~(\ref{upper}), and 
the lower limit at the 68.3\% CL for $\Gamma_h^{\rm min}$ given in Eq.~(\ref{lower}), we find that the 
span of the Higgs width is  
\be
0.71 \leq \frac{\Gamma_h}{\Gamma_h^{\rm SM}} \leq 1.34  \,    ~~.
\ee
Note that this span is not a standard confidence interval, because the lower and upper limits arise from separate measurements. 

The central value of $\Gamma_h^{\rm max}$ is smaller than that of $\Gamma_h^{\rm min}$. This is not a problem 
given that both $\Gamma_h^{\rm max}$ and $\Gamma_h^{\rm min}$ are currently rather broad distributions (see Fig.~1),
so that the $1\sigma$ upper limit on $\Gamma_h^{\rm max}$ is larger than the $1\sigma$ lower limit on $\Gamma_h^{\rm min}$.
It is conceivable, though, that more precise future measurements would yield $\Gamma_h^{\rm max} < \Gamma_h^{\rm min}$
at a confidence level of several standard deviations. The likely interpretation of that situation would be that higher $SU(2)_W$ representations
have VEVs, so that $\Gamma_h^{\rm max}$ is rescaled by $(\cc_V^{\rm max})^4$, with $\cc_V^{\rm max} > 1$.

\section{Limits on Higgs couplings and non-standard decays}\setcounter{equation}{0}

In this section we use the constraints on $\Gamma_h$ obtained in the previous section to set an upper limit on 
the branching fraction $B_X$ into exotic final states, and to derive nearly model-independent ranges (which we call spans) 
for the Higgs couplings.

\subsection{Upper limit  on the exotic branching fraction}

An important implication of the upper limit on $\Gamma_h$ is that it leads to an upper limit on the branching fraction for $h^0$ decays into non-SM final states, ${\cal B}_X$.
Dividing Eq.~(\ref{total-width}) by $\Gamma_h$ gives
\be
{\cal B}_X = 1 - \frac{1}{\Gamma_h} \,  \sum_{\parbox[t]{1.4cm}{\scriptsize $ \mc = W, Z, \\ \hspace*{0.3cm}   b, \tau, g,  \gamma $} } \!\! \cc_\mc^2 \; 
\Gamma^{\rm SM}( h^0 \to \mc\mc )  ~~.
\ee
Using then the upper limit on $\Gamma_h$ given in Eqs.~(\ref{upper-limit}) or (\ref{upper-limit-doublet}), we find the following upper limit  on  the branching fraction into exotic final states:
\be
{\cal B}_X \leq  {\cal B}_X^{\rm max} = 1 -  \left(\frac{\Gamma_h^{\rm SM}}{\Gamma_h^{\rm max} }\right)^{\! 1/2}  \sum_{\parbox[t]{1.4cm}{\scriptsize $\mc = W, Z, \\ \hspace*{0.3cm}   b, \tau, g,  \gamma $} } \!\! 
\ac_\mc \,  {\cal B}^{\rm SM}( h^0 \to\mc\mc )    ~~.
\label{BX-limit}
\ee
In the case of doublet VEVs ($\cc_W = \cc_Z\equiv \cc_V$), the upper limit takes the simpler form
\be
{\cal B}_X^{\rm max} = 1 -  \ac_V  \sum_{\parbox[t]{1.4cm}{\scriptsize $\mc = W, Z, \\ \hspace*{0.3cm}   b, \tau, g,  \gamma $} } \!\! 
\ac_\mc \,  {\cal B}^{\rm SM}( h^0 \to\mc\mc ) 
\label{BX-Vlimit}
\ee

The values of $\ac_\mc$ given in Eqs.~(\ref{aV}) and (\ref{ab})-(\ref{atau}) lead to the following upper limit: 
\be
{\cal B}_X \leq  {\cal B}_X^{\rm max} =  -0.33^{+0.39}_{-0.49} ~~.
\ee
Using the Feldman-Cousins prescription for this manifestly positive observable, we find that the extracted theoretical upper limit is
${\cal B}_X^{\rm max} < 14\%$  at the 68.3\% CL and ${\cal B}_X^{\rm max} < 47\%$  at the 95\% CL.
For precise measurements of several cross sections, the uncertainties in  $\ac_\mc$ may become small enough to turn the upper limit on ${\cal B}_X $
into a severe constraint on physics beyond the SM. 

The above limits are derived under the assumption $C_V^{\rm max} = 1$. For larger values of $C_V^{\rm max}$ the limits are relaxed. 
For example, using $C_V^{\rm max} \approx 1.5$ as in the Georgi-Machacek model  \cite{Georgi:1985nv,Logan:2010en} 
we find ${\cal B}_X^{\rm max} < 59\%$  at the 68.3\% CL and ${\cal B}_X^{\rm max} < 76\%$  at the 95\% CL.

\subsection{Spans of the Higgs couplings}

From Eq.~(\ref{relation}), we find that the various Higgs couplings discussed in Section 2, $\cc_\mc$ for $\mc = W, Z,  b, \tau, g,  \gamma $, can also be bracketed 
between lower and upper bounds extracted from current data:
\be
\ac_\mc^{1/2}  \left( \frac{\Gamma_h^{\rm min}}{\Gamma_h^{\rm SM} }\right)^{\! 1/4} \! < \, \cc_\mc \, < \ac_\mc^{1/2}  \left( \frac{\Gamma_h^{\rm max} }{\Gamma_h^{\rm SM} }\right)^{\! 1/4}   ~~.
\label{lower-upper}
\ee
Using the distributions for $\Gamma_h^{\rm max}$ (see Fig.~1), $\Gamma_h^{\rm min}$ and $\ac_\mc$,  
we find the following 68.3\% (95\%) spans for the Higgs couplings:
\bear
 (0.74) \;\, 0.97 < |\cc_V | \leq 1 \hspace*{1.6cm}  ~~,         &    \hspace*{1.4cm}   &  (0.32) \;\, 0.73 < | \cc_b | < 1.42 \;\, (2.34) ~~,
\nonumber \\ 
 (0.61) \;\, 0.77 < | \cc_g | < 1.07 \;\, (1.63) ~~,      &   &  \hspace*{.4cm}    (0.0) \;\, 0.3 <  | \cc_\tau | < 1.4 \;\, (1.9)  ~~,
\nonumber \\ 
 (0.92) \;\, 1.19 < | \cc_\gamma | < 1.54 \;\, (1.93)  ~~.  & &
 \label{spans}
\eear
These spans arise from the lower limit on  $\ac_\mc^{1/2} (\Gamma_h^{\rm min})^{\! 1/4}$ and the upper limit on  
$\ac_\mc^{1/2} (\Gamma_h^{\rm max})^{\! 1/4}$,  which are obtained from separate experimental inputs, 
so that they should not be interpreted as standard confidence intervals.

\begin{figure}[t]\center
\vspace*{-0.01cm}\hspace*{-0.9cm}
\psfig{file=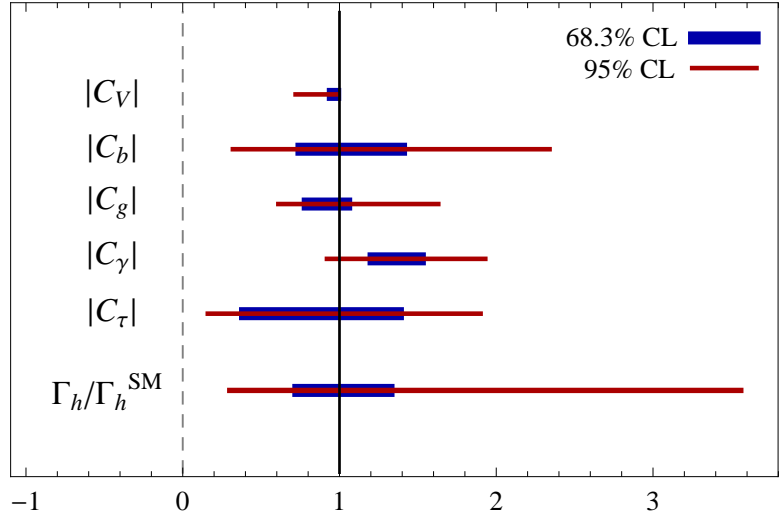 , width=10.9cm,angle=0}
\vspace*{-0.7cm}
\caption{Spans of the Higgs couplings [see Eq.~(\ref{lower-upper})] and total width, normalized to the SM values. The vertical 
lines at 0 and 1 correspond to no Higgs boson, and to the SM, respectively. The left-hand edge of each thick (thin) line  
represents the lower $1\sigma$ (95\% CL) point on the distribution for the lower limit,
while the right-hand edge represents the upper $1\sigma$ (95\% CL) point on the distribution for the upper limit (as shown in Fig.~1 for $\Gamma_h$).
If triplets or higher $SU(2)_W$ representations have VEVs, then the upper limits are pushed to higher values. 
}
\label{fig:CpPlot}
\end{figure}

Note that the upper limit on $\cc_V$ is our input based on the assumption that electroweak symmetry breaking is entirely 
due to the VEVs of doublets.  Using the Feldman-Cousins prescription to take into account this prior, we find that the lower limit on 
 $|\cc_V |$ is relaxed:  $|\cc_V | > 0.93$ (0.72) at the 68.3 (95)\% CL.

The interval for  $\cc_\tau$  is the least reliable, given the large uncertainties discussed 
before Eq.~(\ref{aV}) and after Eq.~(\ref{atau}). Using the Feldman-Cousins prescription for the lower limit $|\cc_\tau| > 0$,
this is shifted to 0.16 at the 95\% CL.
The fact that the SM value of $\cc_\mc =1 $ is within the  95\% span for each of the above five 
couplings is remarkable (see Fig.~2). Nevertheless, new physics contributions may still have effects larger than 50\% on some of these couplings.

Eq.~(\ref{upper-limit}) implies that the upper limits on the $\cc_\mc$ parameters scale as $C_V^{\rm max}$. The values shown in Eq.~(\ref{spans}) correspond to 
$C_V^{\rm max} =1$, while  $C_V^{\rm max} \approx 1.5$ in models with large triplet VEVs \cite{Logan:2010en}. Even larger values of $C_V^{\rm max}$ are 
allowed if scalars transforming as 4 of $SU(2)_W$ acquire VEVs \cite{Low:2010jp}, but models of this type include several charged particles that 
can be ruled out or discovered at the LHC in the near future.

\section{Conclusions}

To determine the true nature of the Higgs-like resonance discovered by the ATLAS and CMS experiments, we need
precise determinations of its underlying couplings to SM particles, extracted without making unnecessary
theoretical assumptions. The impossibility of measuring $\Gamma_h$ at the LHC therefore poses a significant problem,
in addition to masking whether the new resonance has exotic decays that may be difficult to observe experimentally.
We have taken a novel approach to this problem, by observing that $\Gamma_h$ has a model-independent lower
bound, and an upper bound that relies only on the weak assumption that the Higgs-like couplings to $WW$ and $ZZ$ is not
larger than (or not much larger than) the SM values. We showed that $\Gamma_h^{\rm min}$ and
$\Gamma_h^{\rm max}$ can be extracted separately from data for different combinations of Higgs-like signal strengths.
This allows to confine $\Gamma_h$ itself to a certain range between lower and upper limits; this span is not a standard
confidence interval, but the limits themselves, being extracted from data, have 68.3\% and 95\% CL intervals that we
have estimated. It is nontrivial that the resulting span for $\Gamma_h$ is approximately centered on the SM value.

This same information can then be propagated to both an upper limit on the Higgs exotic branching fraction,
and pairs of lower and upper limits for various
Higgs couplings to SM gauge bosons and fermions. For the exotic branching fraction the extracted value of the upper bound
is 14\% at 68.3\% CL in the underlying data, and 47\% at 95\% CL. For the Higgs couplings we find the spans
displayed in Fig.~2. At 95\% CL in the extracted limits all of these spans include the SM value. Notice however
that by far the largest uncertainty in the current extraction of limits applies to the determination of $\Gamma_h^{\rm min}$ and
$\Gamma_h^{\rm max}$ themselves.

With additional data the methodology described here will give increasingly important constraints on the properties
of the newly discovered particle, and is complementary to other approaches currently being pursued.
The major shortcoming of our analysis is our ignorance of the details of the experimental uncertainties
in the published data; this can be overcome easily if the experimental collaborations themselves perform the analysis
that we are advocating.



\vfil
\end{document}